\documentclass[sigconf]{acmart}

\usepackage{booktabs}        
\usepackage{listings}        
\usepackage{booktabs}        
\usepackage{color, colortbl} 
\usepackage{tabulary}        
\usepackage{multirow}        

\renewcommand{\arraystretch}{1.2} 
\newcommand{\ra}[1]{\renewcommand{\arraystretch}{#1}}

\definecolor{LightCyan}{rgb}{0.88,1,1}

\setcopyright{rightsretained}

\acmConference[LLVM-HPC2017]{The Fourth Workshop on the LLVM Compiler Infrastructure in HPC}{November 13th, 2017}{Denver, Colorado USA}
\acmYear{2017}
\copyrightyear{2017}

\newcommand{\HPCTOOLKIT}{\mbox{HPCToolkit}}

\newcommand{\SCOREP}{\mbox{Score-P}}

\graphicspath{{figures/}}

\begin{document}

\copyrightyear{2017}
\acmYear{2017}
\setcopyright{usgovmixed}
\acmConference[LLVM-HPC'17]{LLVM-HPC'17: Fourth Workshop on the LLVM Compiler Infrastructure in HPC}{November 12--17, 2017}{Denver, CO, USA}
\acmBooktitle{LLVM-HPC'17: LLVM-HPC'17: Fourth Workshop on the LLVM Compiler Infrastructure in HPC, November 12--17, 2017, Denver, CO, USA}
\acmPrice{15.00}
\acmDOI{10.1145/3148173.3148187}
\acmISBN{978-1-4503-5565-0/17/11}

\title{An LLVM Instrumentation Plug-in for {\SCOREP}}

\author{Ronny Tsch{\"u}ter}
\affiliation{\institution{Technische Universit\"at Dresden}}
\email{ronny.tschueter@tu-dresden.de}

\author{Johannes Ziegenbalg}
\affiliation{\institution{Technische Universit\"at Dresden}}
\email{johannes.ziegenbalg@tu-dresden.de}

\author{Bert Wesarg}
\affiliation{\institution{Technische Universit\"at Dresden}}
\email{bert.wesarg@tu-dresden.de}

\author{Matthias Weber}
\affiliation{\institution{Technische Universit\"at Dresden}}
\email{matthias.weber@tu-dresden.de}

\author{Christian Herold}
\affiliation{\institution{Technische Universit\"at Dresden}}
\email{christian.herold@tu-dresden.de}

\author{Sebastian D{\"o}bel}
\affiliation{\institution{Technische Universit\"at Dresden}}
\email{sebastian.doebel@tu-dresden.de}

\author{Ronny Brendel}
\affiliation{\institution{Oak Ridge National Laboratory}}
\email{brendelr@ornl.gov}

\renewcommand{\shortauthors}{Tsch{\"u}ter et al.}

\newcommand{\correctme}[1]{\textcolor{red}{#1}}
\newcommand{\stillneeded}[1]{\marginpar{still needed?}\textcolor{cyan}{#1}}
\def \todo{\textbf{\textcolor{yellow}{TODO}}}
\def \citationneeded{\textbf{\textcolor{yellow}{CITATION NEEDED}}}

\begin{abstract}
Reducing application runtime, scaling parallel applications to higher numbers of processes/threads, and porting applications to new hardware architectures are tasks necessary in the software development process.
Therefore, developers have to investigate and understand application runtime behavior.
Tools such as monitoring infrastructures that capture performance relevant data during application execution assist in this task.
The measured data forms the basis for identifying bottlenecks and optimizing the code.

Monitoring infrastructures need mechanisms to record application activities in order to conduct measurements.
Automatic instrumentation of the source code is the preferred method in most application scenarios.
We introduce a plug-in for the LLVM infrastructure that enables automatic source code instrumentation at compile-time.
In contrast to available instrumentation mechanisms in LLVM/Clang, our plug-in can selectively include/exclude individual application functions.
This enables developers to fine-tune the measurement to the required level of detail while avoiding large runtime overheads due to excessive instrumentation.
\end{abstract}
%
%
%
%



\maketitle

\pagebreak
\section{Introduction}
\label{sec:introduction}

\begin{figure}[htb]
	\centering
	\includegraphics[width=0.9\columnwidth]{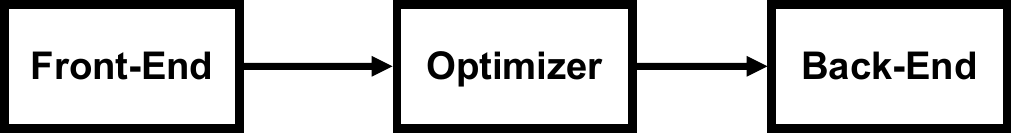}
	\caption{Principle design of a three-phase compiler.}
	\label{fig:three_phase_compiler}
\end{figure}

Compilers are a central component in software development.
They translate statements of a high-level programming language into machine dependent instructions.
In principle, the software design of a compiler consists of the phases: \emph{Front-End}, \emph{Optimizer}, and \emph{Back-End}.
Figure~\ref{fig:three_phase_compiler} illustrates the major components of a three-phase compiler design~\cite{wirth1996compiler}.

\textbf{Front-End}
The Front-End handles code input of a specific source language.
It transforms the source code of the individual language to a generic internal representation (intermediate representation, IR).
Therefore it performs lexical, syntactic, and semantic analyses of the source code.

\textbf{Optimizer}
The code optimization phase transforms the IR while retaining its functionality.
It selectively applies multiple optimization techniques to achieve different goals, e.g., optimize for performance or optimize for size.
Common actions of the Optimizer are register allocation, function inlining, or elimination of unused code segments.

\textbf{Back-End}
The Back-End targets a specific hardware architecture.
It generates machine-dependent code from the transformed intermediate representation.

As already mentioned the Optimizer automatically transforms and optimizes the code.
This allows software developers to easily obtain (performance) benefits by using compiler switches.
However, when it comes to investigating application runtime behavior this technique poses additional hurdles.
As the compiler transparently applies optimizations, it is sometimes unclear how it transforms high-level statements into specific machine code.
For instance, it might not be discernible whether a specific function is inlined by the compiler.
Some compilers provide textual optimization reports listing all applied actions.
However, such reports are typically extensive and cumbersome to interpret manually.
A more intuitive option are debugging and performance analysis tools assisting users in investigating application runtime behavior.
Such tools typically involve a monitoring component observing the application at runtime and recording performance data.
Based on this data, user-friendly graphical visualizations can intuitively depict an application execution or automated analyses may reveal performance bottlenecks.
The monitoring component requires a compiler interface in order to obtain information from the machine code.
This data acquisition should not impede compiler optimizations.
Rather should the obtained data reflect optimizations applied by the compiler.

\begin{figure}
	\centering
	\includegraphics[width=\columnwidth]{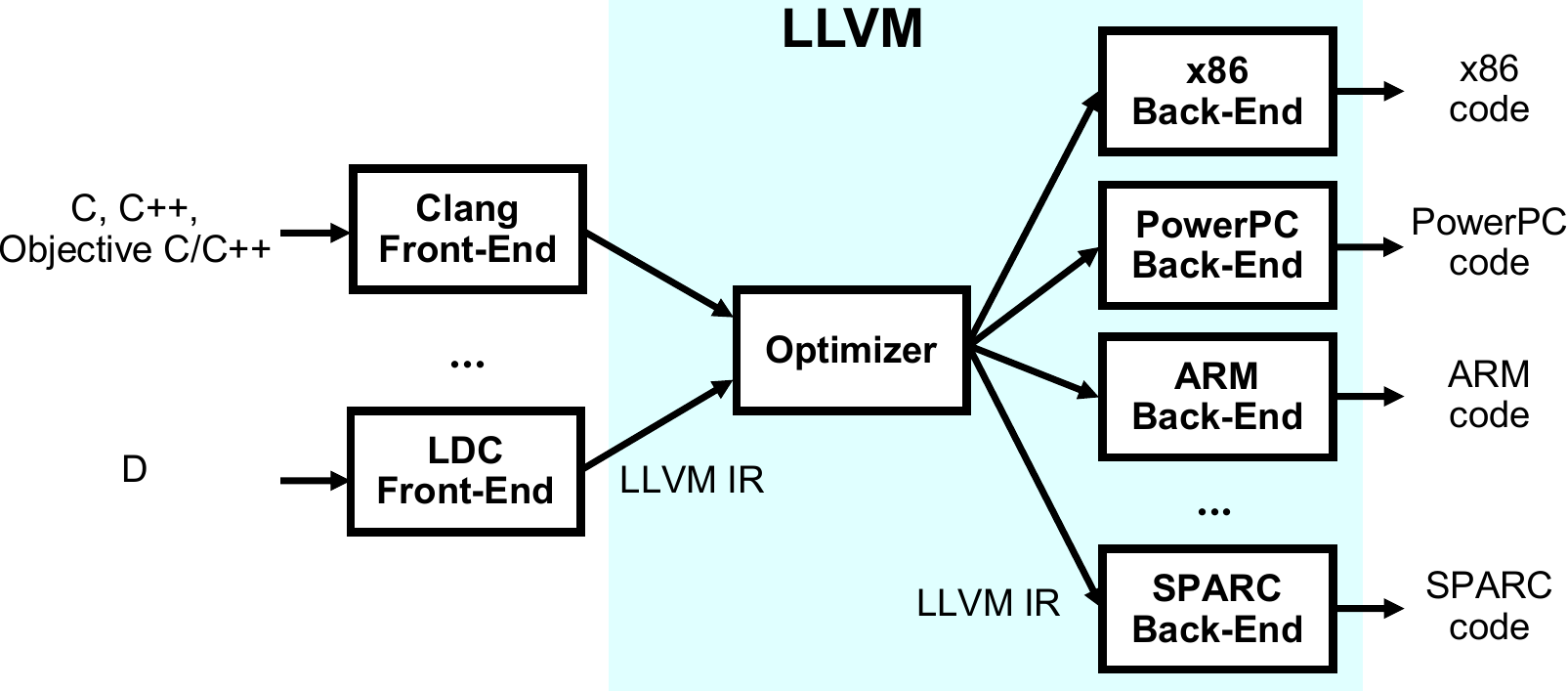}
	\caption{LLVM infrastructure overview.}
	\label{fig:llvm_infrastructure}
\end{figure}

Because of LLVM's growing importance and the fact that it will be one of the main compilers on the upcoming generation of HPC systems at Oak Ridge National Laboratory (Summit~\cite{summit}) and Lawrence Livermore National Laboratory (Sierra~\cite{sierra}), this work focuses on the Low Level Virtual Machine infrastructure (LLVM)~\cite{LLVM:CGO04} and to some extent on the GNU Compiler Collection (GCC)~\cite{gough2004introduction}.
Both are typical representatives of the three-phase compiler design.
Figure~\ref{fig:llvm_infrastructure} depicts the implementation of the three-phase compiler design in the LLVM infrastructure.


In this work we present a new approach for automatic source code instrumentation at compile-time in combination with LLVM.
Our approach records performance relevant data of an application execution for subsequent analyses.
In our work we directly address current limitations of automatic compiler instrumentation in LLVM.
Our specific contributions are:
\begin{itemize}
	\item Implementation of an LLVM pass for automatic source code instrumentation that does not interfere with compiler optimizations
	\item Support for function filtering at
	\begin{itemize}
		\item Compile-time and thereby offering the possibility to avoid the overhead of excessive function instrumentation
		\item Runtime with early evaluation of conditions whether a function is filtered or not directly in the instrumentation code avoiding unnecessary calls into the monitoring component
	\end{itemize}
	\item Support for user defined filter rules
	\item Internal handling of meta data avoiding the need for additional tools in order to obtain, e.g., function names and line numbers
	\item Exception-aware instrumentation improving consistency of recorded data even in case of errors
	\item In-depth performance analysis based on the information provided by our LLVM pass
\end{itemize}


The remainder of this work is organized as follows:
Section~\ref{sec:methodology} presents the design of our LLVM pass for automatic compiler instrumentation.
Section~\ref{sec:implementation} highlights implementation aspects.
Several case studies in Section~\ref{sec:case_study} demonstrate the advantages of our LLVM instrumentation pass.
Section~\ref{sec:related_work} provides an overview of related work.
Lastly, Section~\ref{sec:conlcusion}~and~\ref{sec:future_work} summarize the work and give an outlook on future work.

\section{Methodology}
\label{sec:methodology}

Typically, monitoring infrastructures observe an application at runtime. 
They annotate the source code with so-called \textit{hooks} to record activities like function entries and exits.
The hooks invoke the monitor at each of these events.
This process is also called \emph{instrumentation} and demonstrated in Listings~\ref{lst:original_source_code} and~\ref{lst:annotated_source_code}.
Listing~\ref{lst:original_source_code} shows the unmodified version of a simple code example.
Listing~\ref{lst:annotated_source_code} illustrates the annotated source code of the example.
Each entry and exit of a function now contains a call (hook) to the monitor.
The monitor implements the functions \texttt{ENTER} and \texttt{EXIT}.
Whenever the application enters an annotated function, it calls the hook \texttt{ENTER} and transfers the control flow to the monitor.
The monitor records relevant information such as timestamp or name of the entered function.
Afterwards the monitor returns the control flow back to the application which continues its execution.
The monitor handles the function exit analogously.

\begin{minipage}{0.46\columnwidth}
	\lstset{language=C,
		basicstyle=\ttfamily,
		keywordstyle=\color{blue}\ttfamily,
		stringstyle=\color{red}\ttfamily,
		commentstyle=\color{green}\ttfamily,
		morecomment=[l][\color{magenta}]{\#}}
	\begin{lstlisting}[caption={Original source code},label=lst:original_source_code,frame=tlrb]{Name}
int main()
{
  int i;
 
  for (i=0; i<3; i++)
  {
    func(i);
  }

  return 0;
}
	    
void func(int i)
{

  if (i>0)
  {
    func(i-1);
  }

}
    \end{lstlisting}
\end{minipage}
\hfill
\begin{minipage}{0.46\columnwidth}
	\lstset{language=C,
		basicstyle=\ttfamily,
		keywordstyle=\color{blue}\ttfamily,
		stringstyle=\color{red}\ttfamily,
		commentstyle=\color{green}\ttfamily,
		morecomment=[l][\color{magenta}]{\#},
		moredelim=**[is][\color{red}]{@}{@}}
    \begin{lstlisting}[caption={Annotated source code},label=lst:annotated_source_code,frame=tlrb]{Name}
int main()
{
  int i;
  @ENTER("main");@
  for (i=0; i<3; i++)
  {
    func(i);
  }
  @EXIT("main");@
  return 0;
}

void func(int i)
{
  @ENTER("func");@
  if (i>0)
  {
    func(i-1);
  }
  @EXIT("func");@
}
    \end{lstlisting}
\end{minipage}

Different strategies are possible to insert hooks into source code.
One option is to let users insert hooks manually.
However, manual instrumentation is error-prone, cumbersome, and biased.
Real-world applications with thousands of lines of code render manual instrumentation infeasible.
Consequently, in this work we focus on automatic instrumentation methods.
This section presents two techniques for instrumenting source code with LLVM/Clang.

\paragraph{Automatic Compiler Instrumentation}
This technique uses the \texttt{-finstrument-functions} option of the Clang compiler.
This option instructs the compiler to generate calls instrumenting each function entry and exit.
It works very similar to the analog functionality in the GNU compiler collection.
The compiler annotates function entries with \texttt{\_\_cyg\_profile\_func\_enter} and function exits with \texttt{\_\_cyg\_profile\_func\_exit}.
The monitoring tool needs to implement both functions.
In contrast to the GNU compiler collection, Clang provides no options to exclude functions or source code files from the instrumentation.
The GNU compiler collection provides the flags \texttt{-finstrument-functions-exclude-function-list} and \linebreak \texttt{-finstrument-functions-exclude-file-list} for this purpose.
Therefore, this approach is prohibitive for the instrumentation of short-running, fine-granular functions.
The instrumentation of such functions causes unacceptably large overhead and increases the runtime significantly.
As inserted hooks remain in the binary---and always get called during application execution---even runtime filtering is useless in reducing measurement overhead.
These conditions render automatic compiler instrumentation problematic and raise the need for selective instrumentation at compile-time.

\paragraph{LLVM Compiler Pass}
This technique uses the LLVM Pass Framework.
The LLVM Pass Framework provides an interface to directly modify or interact with a specific task of the compilation process, e.g., the optimization.
This allows to circumvent the drawbacks of the automatic compiler instrumentation technique.
Depending on the goal of the pass, its implementation inherits from one of the following classes:
\begin{itemize}
	\item ModulePass
	\item CallGraphSCCPass
	\item FunctionPass
	\item LoopPass
	\item RegionPass
	\item BasicBlockPass
	\item MachineFunctionPass
\end{itemize}

\begin{figure}
	\centering
	\includegraphics[width=0.9\columnwidth]{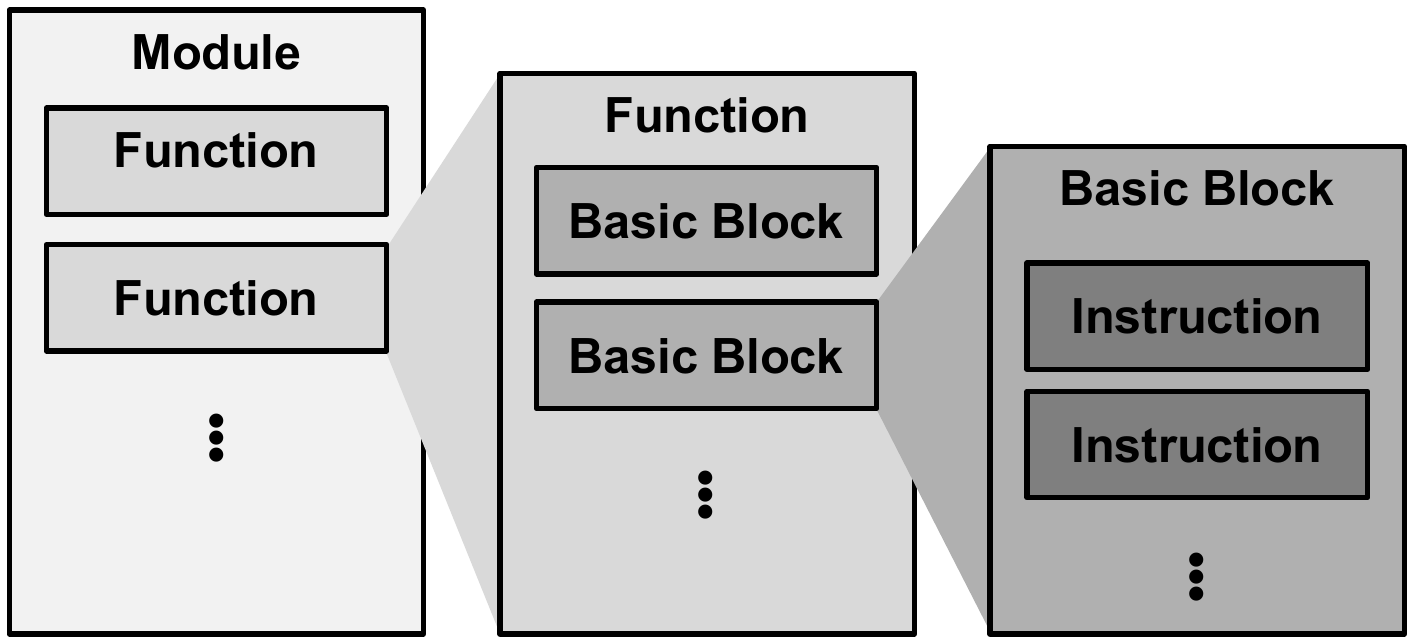}
	\caption{Portion of the LLVM intermediate representation (IR) relevant for this work.}
	\label{fig:module_function_basicblock}
\end{figure}

The classes reflect the entities of the LLVM Intermediate Representation (IR).
Figure~\ref{fig:module_function_basicblock} depicts the concept of the IR.
A module represents the complete application code.
It consists of several functions.
Each function contains multiple basic blocks.
A basic block involves individual instructions.
LLVM provides access to each layer and iterators for selecting individual items.
The class \texttt{FunctionPass} is most suitable for function instrumentation.
It is invoked for each application function.
We use \texttt{FunctionPass} to insert hooks into the IR.
By applying filtering techniques we realize selective function instrumentation at compile-time.
We discuss our implementation in more detail in the following section.

\section{Implementation}
\label{sec:implementation}

\begin{figure}
	\centering
	\includegraphics[width=0.9\columnwidth]{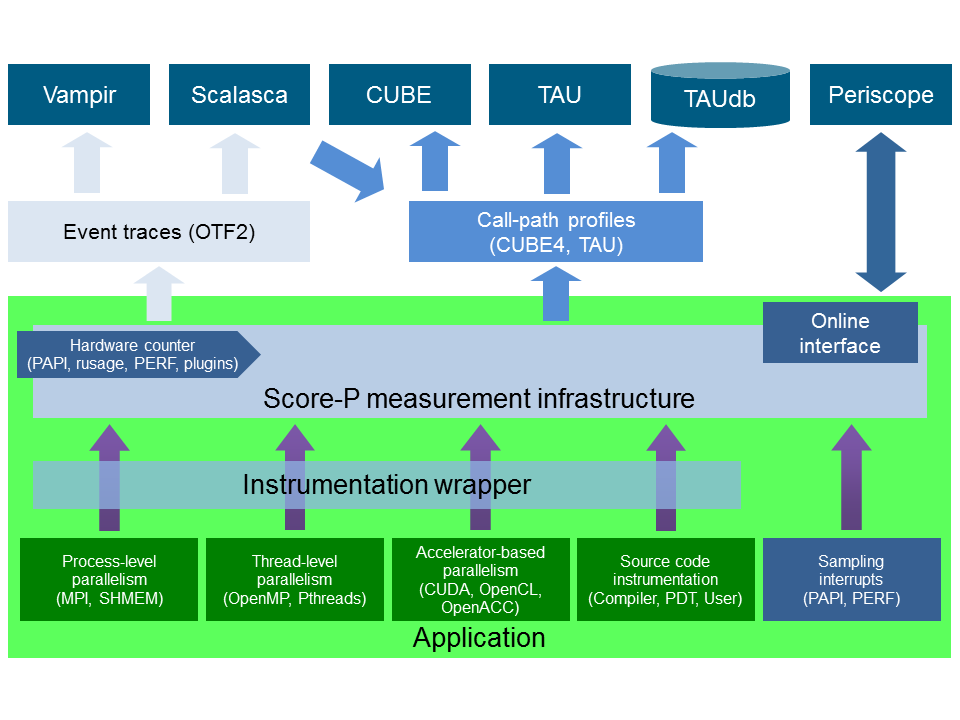}
	\caption{Overview of the Score-P monitoring infrastructure and related analysis tools.}
	\label{fig:scorep_overview}
\end{figure}

In this work we extend the {\SCOREP} monitoring infrastructure~\cite{scorep:2012} to improve its capability for instrumenting code with the Clang compiler.
Figure~\ref{fig:scorep_overview} depicts the general software architecture of {\SCOREP}.
{\SCOREP}~already supports automatic source code instrumentation for GNU, Intel, PGI, Cray, IBM, and Fujitsu compilers.
In addition, {\SCOREP}~provides a GCC compiler instrumentation plug-in~\cite{sp-gcc-plugin} since version 1.4 (released in January 2015).
We have applied the knowledge gained during the GCC plug-in development for the implementation of our LLVM plug-in.

A major requirement of our plug-in is its independence from the programming language of the source code.
This requirement disqualifies an implementation within the Front-End and suggests an LLVM pass implementation.
Our work builds on LLVM version 3.8 or newer.
In addition, we rely on Clang flags to load the dynamic library containing the instrumentation plug-in.
As explained in Section~\ref{sec:methodology} we instrument a source code function by implementing a \texttt{FunctionPass}.
We override the virtual method \texttt{runOnFunction(Function \&F)} inherited from the \texttt{FunctionPass} class.
This \texttt{runOnFunction} method is called for each function in the processed IR.
The parameter \texttt{F} provides the current function.

\paragraph{Collecting meta data}
We utilize the compiler capabilities to collect and store function specific meta data.
For example, we determine the name of the current function.
This is necessary as the compiler might decorate the function name.
For example, in case of function overloading this technique is necessary.
There might exist multiple variants of a function with the same name but different signatures.
In such case, the compiler creates unique internal function names.
Yet, these internal function names are not created considering human readability.
Therefore, we demangle the internal function name to provide function names corresponding to the source code.
In addition, we record source file name and line numbers associated with the current function.
Function names and line numbers allow users to correlate monitored events with the application source code.
By internal meta data handling the plug-in is able to pass this information to the monitoring infrastructure.
Otherwise, the monitoring infrastructure would need external tools or libraries to retrieve this data at runtime.

\paragraph{Deciding whether a function is instrumented}
Within the method \texttt{runOnFunction} we check whether we should instrument the current function.
Therefore, we test certain attributes of the function.
For example, we skip the instrumentation if the function has an empty body, represents a built-in function, or is an internal function of the OpenMP runtime.
Additionally, with our implementation users can selectively remove functions from instrumentation at compile-time.
Thus, we apply a user specified rule set on each function name.
The rule set can contain mangled and demangled function names as well as source file names.
Eventually, considering all checks, we decide whether we filter---exclude from instrumentation---or instrument the current function.

\paragraph{Adding calls to the monitoring infrastructure}
As described in Section~\ref{sec:methodology}, we insert hooks into the IR in order to instrument the current function.
Listing~\ref{lst:instrumentation_pseudo_code} shows the result of adding the instrumentation in pseudo code.
Calls to the monitoring infrastructure are guarded by \texttt{if} statements.
These conditional calls allow an efficient runtime filtering as the condition is evaluated within the instrumentation code.
User defined filter rules determine the condition.
Iterators over basic blocks and their instructions help to find the appropriate place for inserting calls to the {\SCOREP}~monitoring infrastructure.
We insert the instrumentation of the function entry as the last instruction of the first basic block, or before the first call instruction of the first basic block.
The structure of the remaining basic blocks is modified in order to achieve a \texttt{try}/\texttt{finally} semantic as shown in Listing~\ref{lst:instrumentation_pseudo_code}.
The \texttt{try} block encapsulates the instructions of the function body, whereas the corresponding \texttt{finally} block contains the function exit instrumentation.

\lstset{language=C,
	basicstyle=\ttfamily,
	keywordstyle=\color{blue}\ttfamily,
	stringstyle=\color{red}\ttfamily,
	commentstyle=\color{green}\ttfamily,
	morecomment=[l][\color{magenta}]{\#}}
\begin{lstlisting}[float,caption={Pseudo code illustrating the concept of the instrumentation with the LLVM plug-in},label=lst:instrumentation_pseudo_code,frame=tlrb]{Name}
FUNCTION:
static uint32_t handle = INVALID_REGION;
static const region_description descr =
{
  .handle_ref     = &handle,
  .name           = __PRETTY_FUNCTION__,
  .canonical_name = __func__,
  .file           = __FILE__,
  .begin_lno      = input_line,
  .end_lno        = end_lno,
  .flags          = 0
};

if ( handle == INVALID_REGION )
  register_region( &descr );

if ( handle != FILTERED_REGION )
  enter_region( handle );

try
{
  /* FUNCTION BODY */
}
finally
{
  if ( handle != FILTERED_REGION )
    exit_region( handle );
}
\end{lstlisting}

\paragraph{Instrumentation plug-in usage}
The pass is built as a shared library.
The user has to ensure that the compiler loads this shared library to enable instrumentation at compile-time.
Therefore, the Clang compiler requires the following additional arguments:
\lstset{
	basicstyle=\ttfamily,
	columns=flexible,
	breaklines=true,
	breakautoindent=true,
	breakindent=12pt
}
\begin{lstlisting}[escapeinside={@}{@}]
clang -Xclang -load @\newline\postbreak@ -Xclang <instrumenation_pass_library.so> @\newline\postbreak@ -c main.c
\end{lstlisting}
The LLVM pass registry manages registration and initialization of the pass subsystem at compiler startup.\par
After explaining implementation details of our work we demonstrate the usability of the presented instrumentation approach in the following section.

\section{Case Study}
\label{sec:case_study}

In this section we present experiments evaluating the LLVM instrumentation plug-in.
First, we compare event sequences recorded by LLVM's automatic compiler instrumentation and our LLVM instrumentation plug-in.
Second, we demonstrate the usability of our LLVM instrumentation plug-in using a computation kernel of a real world application.
Our test system features $2$ IBM POWER8 CPUs with $10$ cores per CPU ($20$ cores in total). 
We compile the source code of the experiments with Clang 4.0 and GCC 5.4.0.
The experiments use the parallelization paradigms Message Passing Interface (MPI)~\cite{mpi} and OpenMP~\cite{openmp} to distribute the workload over multiple processing elements.
In our experiments we use IBM Spectrum MPI 10.1 and the OpenMP runtime of the corresponding compilers.

\paragraph{Comparison of event sequences}
In the first experiment we compare sequence lengths recorded from a Jacobi solver application.
The Jacobi method implements an iterative algorithm in order to compute the solutions of a diagonally dominant system of linear equations.
This code uses MPI and OpenMP.

In general, the integration of our LLVM instrumentation plug-in into the {\SCOREP} infrastructure provides benefits.
For example, we can utilize already existing features of the {\SCOREP} monitoring infrastructure and corresponding analysis tools.
We instrument the Jacobi solver with {\SCOREP} and subsequently execute the application in order to record its activities at runtime.
We use our LLVM plug-in to instrument user functions.
In addition, {\SCOREP} supports the instrumentation of MPI and OpenMP.
\begin{figure}[hb]
	\centering
	\includegraphics[width=0.9\columnwidth]{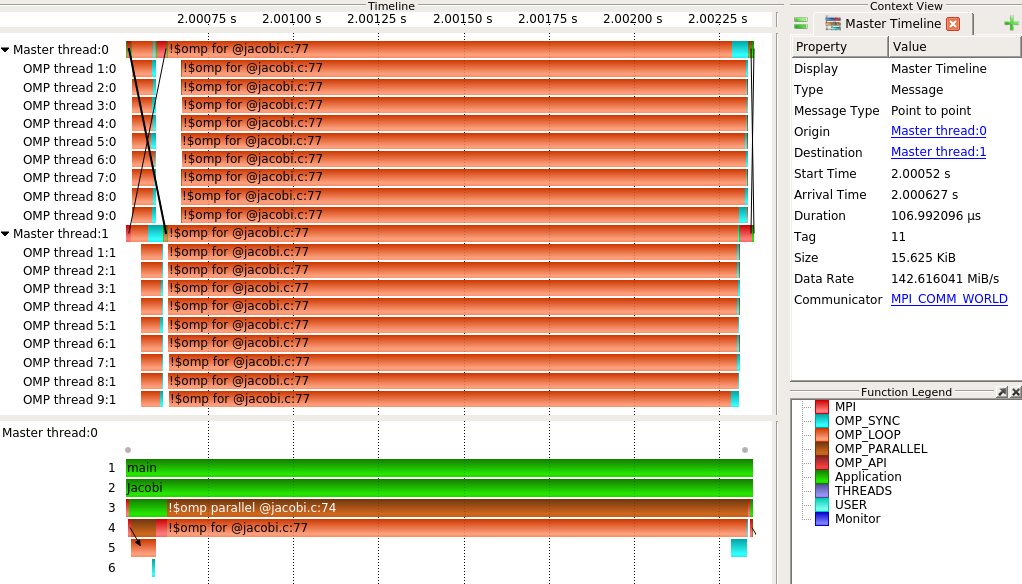}
	\caption{Overview of one Jacobi iteration. Users can investigate user function calls, the usage of parallelization paradigms, and the communication behavior of the application.}
	\label{fig:jacobi_iteration_overview}
\end{figure}
Figure~\ref{fig:jacobi_iteration_overview} shows a timeline visualization of the recorded event sequence in Vampir~\cite{vampir:2008}.
The figure visualizes the application activities in a chronological order.
The top chart shows an overview of all processing elements (represented by horizontal bars) including their executed activities (represented by individual colors in the bars).
It depicts two MPI processes, ``Master thread:0'' and ``Master thread:1''.
Each MPI process forks nine additional threads in OpenMP parallel regions (``OpenMP thread'').
The bottom chart visualizes the call stack of ``Master thread:0''.
Green areas indicate invocations of user functions\footnote{\emph{User functions} are all functions belonging to the application source code excluding calls to libraries and parallelization paradigms.}.
Brown areas in the figure indicate OpenMP parallel regions, whereas orange areas indicate OpenMP parallel loops.
Red areas represent calls to MPI library functions.
Black lines between the MPI processes depict message transfers.
As shown in the right side of Figure~\ref{fig:jacobi_iteration_overview}, users can access information about parallelization paradigms in addition to source code instrumentation.
This is vital for comprehensive performance analyses since performance bottlenecks are often caused by complex effects.
To improve the scalability of parallel applications, programmers have to investigate the interactions between user code, library calls, and parallelization paradigms.

In the next experiment we compile the source code with Clang using different optimization levels---\texttt{-O0} to \texttt{-O3}---and both instrumentation options, \texttt{-finstrument-functions} and our LLVM instrumentation plug-in.
The recorded event sequence---representing function enters and exits---differs between both instrumentation methods.
Table~\ref{tab:jacobi_function_invocations} compares the experiment variants and their corresponding number of user function invocations summarized over all processing elements.
\begin{table}\centering
	\ra{1.3}
	\caption{Number of user function invocations over all processing elements.}
	\label{tab:jacobi_function_invocations}
	\begin{tabular}{m{0.2\columnwidth}m{0.3\columnwidth}m{0.3\columnwidth}}
		\toprule
		& \multicolumn{2}{c}{Number of user function invocations}                   \\
		Optimization level    & automatic compiler instrumentation           & instrumentation via plug-in \\
		\midrule
		\texttt{-O0}          & $2014$                                         & $2014$  \\
		\rowcolor{LightCyan}
		\texttt{-O1}          & $2014$                                         & $2014$  \\
		\texttt{-O2}          & $2014$                                         & $2010$  \\
		\rowcolor{LightCyan}
		\texttt{-O3}          & $2014$                                         & $2008$  \\
		\bottomrule
	\end{tabular}
\end{table}

\begin{figure}[ht]
	\centering
	\includegraphics[width=0.9\columnwidth]{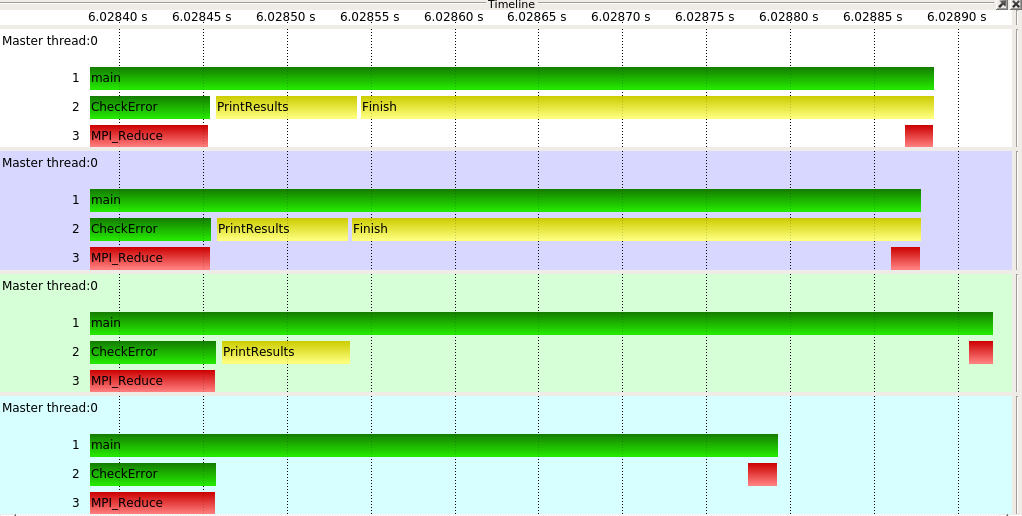}
	\caption{Call stack visualization of the Jacobi application compiled with different optimization levels. The recorded event sequence differs for the optimization levels \texttt{-O0} (white background), \texttt{-O1} (blue background), \texttt{-O2} (green background), \texttt{-O3} (azure background). Yellow areas indicate functions that are inlined in higher optimization levels.}
	\label{fig:jacobi_callstack_comparison}
\end{figure}

\begin{sloppypar}
Automatic compiler instrumentation using \texttt{-finstrument-functions} produces the same event sequence lengths for all optimization levels, comprising of $2014$ function calls in total.
The LLVM instrumentation plug-in generates event sequences containing less function calls in higher optimization levels.
The interaction between optimizations performed by the compiler and the selected instrumentation method causes this different behavior.
For instance, the handling of function inlining and function instrumentation depends on the compiler.
The automatic compiler instrumentation of GCC also annotates functions expanded inline in other functions~\footnote{\url{https://gcc.gnu.org/onlinedocs/gcc/Instrumentation-Options.html}}.
In contrast, the usage of \texttt{-finstrument-functions} disables function inlining for the Intel C/C++ compiler~\footnote{\url{https://software.intel.com/en-us/node/682535}}.
Although it is not clearly documented, we expect that LLVM's automatic compiler instrumentation annotates inlined functions similar to GCC.
Nevertheless, the instrumentation pass of our LLVM plug-in runs after the compiler optimization.
If a function is inlined, it does not represent a function call.
Consequently, inlined functions are not instrumented by the plug-in and do not trigger an event at runtime.
Figure~\ref{fig:jacobi_callstack_comparison} contrasts event streams recorded with different optimization levels.
It illustrates the call stack of the master thread of the first MPI process of the Jacobi application.
The comparison includes results for the optimization level \texttt{-O0} (white background), \texttt{-O1} (blue background), \texttt{-O2} (green background), and \texttt{-O3} (azure background).
Yellow bars highlight functions that are inlined in higher optimization levels.
\end{sloppypar}

\paragraph{Comparison of runtime overheads}
In the second experiment we investigate instrumentation inflicted runtime overheads using the application miniFE~\footnote{\url{http://www.nersc.gov/users/computational-systems/cori/nersc-8-procurement/trinity-nersc-8-rfp/nersc-8-trinity-benchmarks/minife/}}.
miniFE is an application of the Trinity benchmark suite and consists of a couple of compute kernels representing implicit finite-element applications.
The code consists of an OpenMP and a MPI+OpenMP version.
In our experiments we use the OpenMP version of miniFE.
Although {\SCOREP}~can record the parallelization paradigm (see the Jacobi example above), we will focus on source code instrumentation in the following.
Table~\ref{tab:minife_runtime} shows the minimum runtime of three respective experiment runs.
\begin{table}\centering
	\ra{1.3}
	\caption{Runtime in seconds of the miniFE experiments. Each experiment was executed three times, the minimum of these runs is shown.}
	\label{tab:minife_runtime}
	\begin{tabulary}{0.95\columnwidth}{LL}
		\toprule
		Experiment & Runtime in seconds \\
		\midrule
		uninstrumented & 6 \\  
		\rowcolor{LightCyan}
		automatic compiler instrumentation & $800$ \\  
		automatic compiler instrumentation, runtime filter & $140$ \\  
		\rowcolor{LightCyan}
		instrumentation via plug-in & $27$ \\  
		instrumentation via plug-in, compile-time filter & $7$ \\
		\bottomrule
	\end{tabulary}
\end{table}

\subparagraph{\textit{Uninstrumented baseline:}}
We start with a baseline measurement of the uninstrumented binary.
The miniFE binary expects $3$ input parameters which specify the dimensionality of the problem.
If not stated otherwise we run the experiments with a problem size of $128$ for each of the $3$ dimensions.
The uninstrumented application runs about $6$ seconds using $20$ threads.

\subparagraph{\textit{Automatic compiler instrumentation via \texttt{-finstrument-functions}:}}
\begin{sloppypar}
In the next experiment we instrument miniFE with the \texttt{-finstrument-functions} option of the Clang compiler.
Automatic compiler instrumentation increases the runtime by more than a factor of $100$.
A profile of the recorded events contains about $16.9$ billion function calls.
Most of these functions---$16.4$ billion or $97\%$---are C++ STL functions.
The C++ STL functions are executed very frequently whilst showing a short runtime per call.
The instrumentation of this kind of functions induces a high runtime overhead and thereby severely disturbs the application behavior, rendering the recorded event data---especially the timing information---useless.
\end{sloppypar}

\subparagraph{\textit{Automatic compiler instrumentation via \texttt{-finstrument-functions} with runtime filtering:}}
In this experiment setup we want to reduce the overhead by applying a filter at runtime.
This filter prevents the recording of C++ STL events and some miniFE functions handling internal data structures.
However, each instrumented event still calls into the monitoring infrastructure.
With the applied filter the runtime decreases to $140$ seconds, i.e., in comparison to the uninstrumented application the runtime increases by a factor of $23$.
These results show that runtime filtering is not sufficient in order to reduce the overhead to an acceptable amount.
Consequently, we must already avoid the instrumentation of high frequency functions.

\subparagraph{\textit{Instrumentation via LLVM plug-in:}}
In the next experiment we instrument miniFE using our LLVM plug-in.
The instrumentation pass of our plug-in runs after the optimization step.
As a result, the plug-in does not instrument inlined functions, which prevents the annotation of C++ STL functions and thereby dramatically reduces the number of calls into the monitoring infrastructure at runtime.
This setup reduces the runtime to $27$ seconds.
That corresponds to a runtime increase by a factor of $4$ compared to the uninstrumented version.

\subparagraph{\textit{Instrumentation via LLVM plug-in with compile-time filtering:}}
In addition to the previous experiment, we also apply a filter file containing a list of miniFE functions that are called with a high frequency and show a runtime less than $1 \mu$s per visit.
The LLVM instrumentation plug-in uses these filter rules to exclude the specified functions from instrumentation.
With this compile-time filtering we are able to reduce the application runtime to $7$ seconds, which is close to the initial baseline.

\subparagraph{\textit{Summary:}}
In our experiments automatic compiler instrumentation exhibits severe runtime overhead, limiting its usability, especially for C++ applications.
We run additional tests with an increased problem size of $512 \times 256 \times 256$ to evaluate the trend for extended application runs.
In this setup the uninstrumented binary completes the computation in about $90$ seconds.
However, with automatic compiler instrumentation enabled the miniFE application does not finish within $2$ hours.
Using the LLVM instrumentation plug-in we are able to run the application in $310$ seconds.
These measurements support the results presented above and confirm the need for methods that can prevent instrumentation of specific functions.
Additional tests (not shown) with GCC's automatic compiler instrumentation and the GCC instrumentation plug-in in {\SCOREP}~show similar behavior.

\section{Related Work}
\label{sec:related_work}

All authors of this paper work in the area of high performance computing (HPC).
Although our approach is not limited to HPC, we focus on related work in the area of HPC.

\paragraph{Data Acquisition}
Typical HPC machines provide installations of at least one of the commercial Intel~\cite{icc}, PGI~\cite{pgcc}, Cray~\cite{craycc}, IBM~\cite{ibmxl}, and Fujitsu~\cite{fujitsucc} compilers.
However, the Open Source compilers GCC~\cite{gough2004introduction} and LLVM~\cite{LLVM:CGO04} are also often used.
All of them provide compiler options to enable automatic function instrumentation.
For example, GCC, Intel, and LLVM provide the \texttt{-finstrument-functions} option.
On the one hand, the automatic function instrumentation might result in less aggressive code optimization, e.g., inlining will be disabled.
On the other hand, selective function instrumentation is hard to realize with this option.
For example, GCC provides additional options to specify filter rules to exclude specific functions or complete source code from instrumentation.
However, these filter rules work on substring matching.
The filter rule \textit{EXCLUDE foo} does not only prevent instrumentation of function \textit{foo} but also matches with the functions \textit{foobar}, \textit{myfoo}, and \textit{another\_foo\_func}.
LLVM and GCC provide a plug-in interface to implement extension for the compiler.
These plug-in interfaces are used in this work to enhance function instrumentation capabilities.
Since version 4.0 LLVM/Clang provides an additional instrumentation option---the XRay function call tracing system~\cite{xray}.
It implements a MachineFunctionPass.
In contrast to our approach XRay uses binary instrumentation and thereby works closer to the machine level.
It supports filtering of functions at compile-time based on their number of instructions.
Future releases may contain features for advanced filtering options.
Exceptions are not handled gracefully by the XRay instrumentation.

An alternative method to record information about the application runtime behavior is sampling.
Sampling periodically interrupts an application and records its status, e.g., the current call stack.
In contrast to instrumentation---where each instrumented event is triggered at runtime---sampling gives a statistical overview on the application depending on the sampling frequency.

\paragraph{Data Analysis}
Performance and debugging tools are the prime users of information about applications' runtime behavior.
TAU~\cite{tau} can instrument Fortran, C, and C++ applications.
It supports source-level (manipulating the source code), binary-rewriting as well as compiler instrumentation.
Fine-grained selective instrumentation is only available for source-level and binary-rewriting instrumentation.
In addition, TAU provides a feature that avoids recording functions based on the number of their calls and their runtime per call.
This feature can be used to reduce the overhead induced by excessive recording of short-running functions.
However, the feature just skips the data recording but does not avoid the function instrumentation itself.
{\SCOREP}~\cite{scorep:2012} is a common monitoring infrastructure for several analysis tools such as Vampir~\cite{vampir:2008}, Scalasca~\cite{scalasca}, Periscope~\cite{gerndt2010automatic}, and TAU~\cite{tau}.
It supports automatic compiler instrumentation in combination with various compilers.
In addition, {\SCOREP}~is able to instrument functions via the plug-in interface of the GCC compiler since version 1.4.
Due to the standardized data formats a wide range of tools is available to analyze performance data recorded by {\SCOREP}.
Many tools, e.g., Allinea~DDT/MAP~\cite{allineamap}, Intel~VTune~\cite{intelvtune}, Cray~PAT~\cite{kaufmann2003craypat}, \HPCTOOLKIT~\cite{hpctoolkit}, and Extrae~\cite{extrae}, combine instrumentation and sampling.
For example, calls to the MPI library are instrumented whereas user functions are captured via sampling.
GProf~\cite{graham82} instruments in order to get an exact call count for each function, and samples the application to get statistical timing information.
However, these tools store their performance data in individual formats.
As a result, the interoperability of these tools is limited.

\section{Conclusion}
\label{sec:conlcusion}

Common instrumentation approaches employ runtime filtering methods of selected functions in order to control performance data size.
Our plug-in additionally allows for selective instrumentation of specific functions.
This completely avoids the measurement overhead of filtered functions.
Especially for C++ codes, selective instrumentation enables measurement runs with acceptable overhead, making highly-detailed measurements for such applications possible. 

The LLVM compiler infrastructure's clean separation into the Front-End and Optimizer assists our work on the plug-in implementation.
We concentrate on efficient code instrumentation mechanism without having to take care of specific characteristics of the processed programming language.
However, this separation has shortcomings when it comes to transferring additional information from the Front-End to the Optimizer.
First, the Front-End must be instructed to emit necessary information for proper source code location annotation of the instrumented functions (e.g, source file name and line number).
For the Clang Front-End, that means users have to compile their source code with debug information (\texttt{-g}), even if the final object file does not need debug symbols at all.
In our previous work on the GCC instrumentation plug-in we were able to directly obtain such information from GCC.
Second, the IR representation only provides the mangled symbol name for functions, which are hard to read by users. 
However, creating the demangled name requires the use of external tools.
A last shortcoming is the missing annotation of \emph{artificially} created functions, e.g., \texttt{.omp\_outlined.} functions created by the compiler for OpenMP parallel regions or tasks.
In order to identify such functions, we match the name of the currently processed function with a list of known \emph{artificial} functions.
However, these names might be subject to change due to internal refactoring within the LLVM infrastructure. 
Some of these shortcomings could be circumvented by implementing an additional Clang Front-End plug-in.
Though, a Front-End plug-in goes hand in hand with dependencies to specific programming languages and thereby would limit the versatility of our instrumentation plug-in.
Therefore, it is desirable if LLVM provides options to query the mentioned kind of information conveniently.

\section{Future Work}
\label{sec:future_work}

%
In contrast to Clang---the LLVM C/C++ Front-End---there is currently no native Fortran Front-End available for LLVM.
Nevertheless, Fortran is still an important programming language in scientific and high performance computing.
One option for compiling Fortran code with the LLVM infrastructure is DragonEgg~\cite{llvm2013dragonegg}~\footnote{\url{http://dragonegg.llvm.org/}}.
It acts as a GCC plug-in.
Thus, DragonEgg allows the use of GCC's Front-Ends with LLVM's Optimizers and code generators.
The Flang project~\footnote{\url{https://github.com/flang-compiler/flang}} works on the implementation of a native Fortran Front-End for LLVM.
When the Flang Front-End becomes available, we will repeat our experiments.
As our instrumentation plug-in works on the LLVM IR and therefore is independent of a specific Front-End, we do not expect the need for major adjustments in the plug-in code with respect to the instrumentation mechanism.
However, we expect to modify aspects of the meta data handling.
For example, Fortran compilers use custom mangling schemes for Fortran modules which implies adaption to the creation of pretty function names in our plug-in.


\begin{acks}
This research used resources of the Oak Ridge Leadership Computing Facility at Oak Ridge National Laboratory, which is supported by the Office of Science of the Department of Energy under Contract DE-AC05-00OR22725.
\end{acks}

\bibliographystyle{ACM-Reference-Format}
\bibliography{bibliography}

\end{document}